\documentclass[letter]{aa}
\usepackage{graphicx}
\usepackage{txfonts}
\def\spose#1{\hbox to 0pt{#1\hss}}
\def\simlt{\mathrel{\spose{\lower 3pt\hbox{$\mathchar"218$}}
     \raise 2.0pt\hbox{$\mathchar"13C$}}}
\def\simgt{\mathrel{\spose{\lower 3pt\hbox{$\mathchar"218$}}
     \raise 2.0pt\hbox{$\mathchar"13E$}}}

\begin{document}
   \title{The early evolution of tidal dwarf galaxies}

   \author{S. Recchi
          \inst{1, 2}
          \and
          C. Theis\inst{1}
          \and
          P. Kroupa\inst{3}
          \and
          G. Hensler\inst{1}
          }

   \offprints{S. Recchi}

   \institute{Institute of Astronomy, Vienna University,
              T\"urkenschanzstrasse 17, A-1180 Vienna, Austria\\
              \email{recchi, theis, hensler@astro.univie.ac.at}
		\and
              INAF - Osservatorio Astronomico di Trieste, 
              Via G.B. Tiepolo 11, 34131 Trieste, Italy\\
	      \email{recchi@oats.inaf.it}
                \and
              Argelander Institute for Astronomy, Bonn University, 
              Auf dem H\"ugel 71, 53121 Bonn, Germany\\
              \email{pavel@astro.uni-bonn.de}
             }

   \date{Received; accepted}

  \abstract
   {Dwarf galaxies can arise from self-gravitating structures emerging 
        from tidal tails. What fraction of the known dwarf galaxies in the 
        Local Universe can have this origin is still a matter of debate.}
   {In our effort to understand the origin and evolution of tidal dwarf 
	galaxies and their correspondence with local objects, the first 
	step is to understand how these galaxies (which are supposed 
	to have a limited amount of dark matter) react to the feedback of 
	the ongoing star formation.}
   {We make use of 2-D chemodynamical calculations in order to study the 
	early evolution of isolated, dark matter-free dwarf galaxies.  We 
	present models in which feedback parameters are varied.  We 
	also compare the results with dark matter-dominated dwarf galaxy 
        models.}
   {All the considered models show that the star formation proceeds for more 
	than 300 Myr, therefore dwarf galaxies without large dark matter 
        halos are not necessarily quickly destroyed.  The chemical 
	evolution of these objects is consistent with the main chemical 
	properties of the dSphs of the Local Group.  Models with 
	large dark matter halos show results consistent with models 
	free of dark matter, indicating that the distribution of gas is 
	more important than the depth of the potential well in determining 
	the global behaviour of dSph-sized dwarf galaxies.}
    {}


   \maketitle
%

\section{Introduction}

Galactic collisions and close encounters are thought to be common
events in the Local Universe.  It has been clear for some time that
the structure of a galaxy can be affected by tidal interactions with
close neighbours (Toomre \& Toomre \cite{tt72}).  Even for what
concerns the Milky Way, signs of the interaction with the Magellanic
Clouds have been analysed in detail (e.g.  Mastropietro et
al.\ \cite{mast05}; R\accent23u\v{z}i\v{c}ka et al.\ \cite{rpt07}).

There is growing evidence that tidal arms, produced by the interaction
of gas-rich galaxies, can trigger the formation of self-gravitating
structures evolving into objects similar to the local dwarf galaxies
(Duc \& Mirabel \cite{dm98}; Walter et al. \cite{wmo06}).  These
objects are therefore called {\it tidal dwarf galaxies} (TDGs) and
their production rate must have been higher in the early universe than
today, due to the higher probability of tidal interactions and the
larger gas content of young galaxies.  From a stellar dynamical point
of view, it is shown that these star complexes typically survive the
first few hundred Myr (Kroupa \cite{krou98}) and can evolve into
objects similar to the present-day dwarf spheroidal (dSph) galaxies
(Kroupa \cite{krou97}; Metz \& Kroupa \cite{mk07}).  Recently,
detailed numerical simulations of the evolution of tidal tails have
been presented (Bournaud \& Duc \cite{bd06}; Wetzstein et
al. \cite{wnb07}) showing that the formation of TDGs in these sites is
favoured by the presence of large and extended reservoirs of gas.
These studies also show that TDGs must have a very reduced dark matter
(DM) content, confirming previous results (e.g. Barnes \& Hernquist
\cite{bh92}).  Okazaki \& Taniguchi (\cite{ot00}) even point
out that the majority, if not all the dwarf galaxies could be TDGs if
only a few long-lived TDGs emanate from each late-type encounter.

It is therefore important to understand the early evolution of TDGs,
in order to clarify how the potentially disruptive feedback from the
ongoing star formation (SF) acts on a DM-poor structure, characterised
by a very reduced potential well, and to compare them with DM-rich
models.  Finally, the study of the chemical evolution of dwarf
galaxies and comparison with observed ones offers a wealth of
information about the past history and the origin of these objects.

In this paper we study for the first time the 2-D chemodynamical
evolution of DM-free model galaxies, in order to understand under
which conditions these objects can sustain the energy released by the
dying stars without experiencing a complete blow-away, as well as what
the final chemical patterns are.  This work is an extension of the 1-D
chemodynamical simulations of DM-free dwarf galaxies of Hensler et
al. (\cite{htg04}).  In Sect.~\ref{mod} we describe the adopted model
and in Sect.~\ref{res} we summarise the main results of our
investigations.  Finally, in Sect.~\ref{conc} we draw some
conclusions.


\section{The model}
\label{mod}

We performed 2-D simulations of the dynamical and chemical evolution
of TDGs in cylindrical coordinates.  The hydro solver and the
implementation of the routines for the chemical evolution have been
described in Recchi et al. (\cite{rmd01}; \cite{rmdt04}) and
references therein.  Several new implementations have been added in
this version and will be comprehensively described in Recchi et
al. (in preparation).  The main code extensions are:

\begin{itemize}

\item self-gravity like in Rieschick \& Hensler (\cite{rh03}), which 
solves the Poisson equation by means of the so-called ``Alternate
Direction Implicit'' method.

\item Metal-dependent stellar wind luminosities taken from the 
Starburst99 (Leitherer et al. \cite{leit99}) software package.

\item A star formation rate proportional, within each grid cell, 
to the amount of gas (through a constant $\varepsilon_{\rm SF}$) with
an upper temperature threshold (T$_{\rm thr}$) and valid only for
convergent gas flows (i.e. when $\bf \nabla$ $\cdot$ $\bf v$ $<$ 0).
$\varepsilon_{\rm SF}$ denotes the fraction of gas converted to stars
within the next 5 Myrs, once the SF criteria are fulfilled.


\end{itemize}

The calculation strategy is that: the SF criteria identify the grid
points that, within an interval of 5 Myr, produce stars.  The star
masses are distributed according to a Salpeter IMF.  All the relevant
variables concerning this SF generation are stored in arrays that are
used to calculate the feedback from SNeIa, SNeII, and
intermediate-mass stars.  In this way, the chemical composition of
stars is always recorded and metal-dependent yields and energy
production can be taken into consideration properly.  In spite of some
tricks to save memory, the calculations are very computationally
demanding (several weeks to months for each run), therefore only a
limited number of runs can be performed.  For our convenience, the
dynamics of the stars is neglected in this set of models, therefore
the stars remain where they are born.  Due to their short lifetimes,
this is not critical at all for the massive stars, which are the main
sources of stellar energy feedback.

Although these simulations are not aimed at reproducing specific
objects, the Local Group dwarf galaxy Fornax gives us a good
reference for calibrating the initial setup.  Indeed, it has been
suggested that the dSphs near to the Milky Way may be TDGs because
their spatial distribution is incompatible with a cosmological origin
(Kroupa et al. \cite{ktb05}).  Although Local Group dSphs are commonly
believed to be DM-dominated, there are arguments suggesting that they
are not necessarily DM-rich.  Twisted and squashed isophotes are
common in Local Group dSphs and are not compatible with the presence
of a massive DM halo.  Most of the kinematical properties of these
objects can also be explained without DM (Metz \& Kroupa
\cite{mk07}).  An extensive discussion about this point will be
presented in Recchi et al. (in preparation).  The initial gas
distribution is spherically symmetric and follows a King profile with
a core radius of 500 pc. The total gas mass {\rm in the grid (a sphere
of $\sim$ 8 kpc of radius)} is calculated according to M$_g =$ M$_{\rm
Fornax}$ / $\varepsilon_{\rm SF}$, M$_{\rm Fornax}$ being the mass of
Fornax (assumed to be 6.8 $\cdot$ 10$^7$ M$_\odot$, Mateo
\cite{mate98}).

We consider models in which the SF efficiency $\varepsilon_{\rm SF}$
and the temperature threshold T$_{\rm thr}$ are varied.  Two models
with a large rigid DM halo are considered, too, in order to outline
the differences with DM-free models.  The central DM density is
calculated according to the observed velocity dispersion of Fornax
(10.5 km s$^{-1}$, Mateo \cite{mate98}) and the distribution follows a
quasi-isothermal profile with core radii of 1 kpc and 200 pc,
respectively.  These models contain 1.1 $\cdot$ 10$^7$ and 3.7 $\cdot$
10$^7$ M$_\odot$ of DM within 1 kpc, respectively.  We also consider a
model in which the feedback from stellar winds is suppressed.  Model
parameters are summarised in Table ~\ref{tmod}.  A wider range of
values of $\varepsilon_{\rm SF}$ and T$_{\rm thr}$ have been
considered, as well as models with better spatial resolution and
models with an initial anisotropic distribution of gas.  For the sake
of simplicity, we focus in this letter only on the 6 basic models,
mentioning briefly the behaviour of the other models.  Only the first
300 Myr of the evolution have been calculated.  This is due to the
large computational time required by the simulations, but also to the
fact that for timescales longer than this, interaction effects with
surrounding galaxies are supposed to play a major role in the
evolution of these small objects.  The effect of tidal interactions on
the evolution of these objects is currently under investigation.

\begin{table}[ht]
\caption{Model parameters}
\label{tmod}
\begin{center}
\begin{tabular}{ccccc}
\hline \hline
Model & DM? (core radius) & $\varepsilon_{\rm SF}$ & T$_{\rm thr}$ (K) 
& Stellar winds?\\
\hline                        
   1   & No           & 0.1   &  10$^4$            & Yes\\      
   1dc & Yes (200 pc) & 0.1   &  10$^4$            & Yes\\      
   1de & Yes (1 kpc)  & 0.1   &  10$^4$            & Yes\\      
   2   & No           & 0.1   &  5 $\cdot$ 10$^3$  & Yes\\
   3   & No           & 0.1   &  10$^4$            & No\\      
   4   & No           & 0.05  &  10$^4$            & Yes\\      
\hline
\noindent
\end{tabular}
\end{center}
\end{table}

\vspace{-1.2cm}
\section{Results}
\label{res}

\subsection{A typical evolutionary sequence}
\label{res_ref}
   
We show here an example of the chemodynamical evolution of an
isolated, DM-free dwarf galaxy.  This model (designated as Model 1,
see Table ~\ref{tmod}) is characterised by $\varepsilon_{\rm SF} =
0.1$ and T$_{\rm thr} = 10^4$ K.  This temperature is much higher than
the typical temperatures of the cores of molecular clouds, where SF
takes place.  Indeed the size of each computational cell is also much
larger than a molecular cloud core.  (The size of the first cell is 10
pc, then it proceeds radially outwards with a size ratio between
adjacent zones of 1.028.)  Since this temperature has to be considered
as the average temperature of a region of the galaxy encompassing a
star-forming molecular cloud, T$_{\rm thr}$ is necessarily a free
parameter in our models.  We discuss its effect in
Sect.~\ref{res_others}.

\begin{figure}[ht]
 \begin{flushleft}
 \includegraphics[width=9cm]{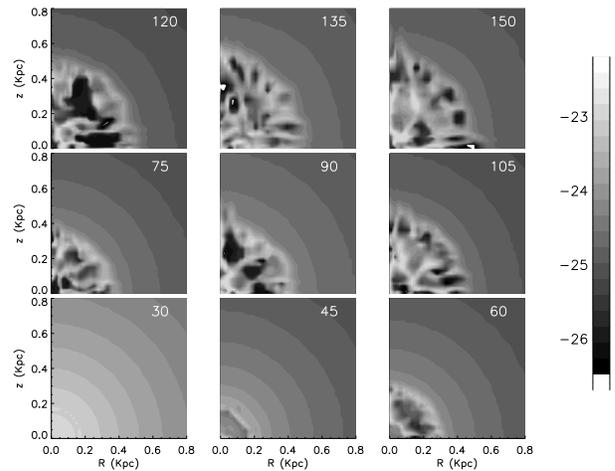}
\caption{ Density contours of the gas for Model 1 at 9 evolutionary 
times (labelled, in Myr, at the top right corner of each panel). The
density scale (in g cm$^{-3}$) is on the righthand strip.  }
\label{9s} 
\end{flushleft}
\end{figure}

\begin{figure}[ht]
 \begin{flushleft}
 \vspace{-0.5cm}
 \hspace {0.4cm} \includegraphics[width=8cm]{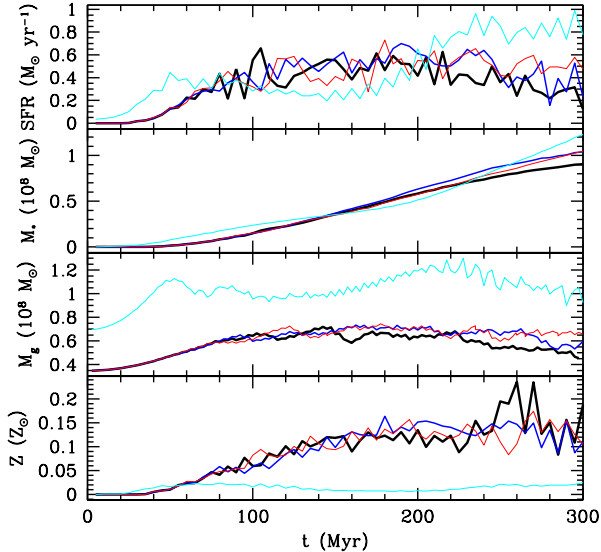}
\caption{ Evolution with time of the SFR (in M$_\odot$ / 
yr, upper panel), stellar mass (in 10$^8$ M$_\odot$, second panel),
gas mass within 1 kpc (in 10$^8$ M$_\odot$, third panel), and mean
metallicity of the stellar component (in Z$_\odot$, lower panel) for
Model 1 (reference model, very thick lines), Model 2 (reduced T$_{\rm
thr}$, thick lines), Model 3 (no stellar winds, thin lines), and Model
4 (reduced $\varepsilon_{\rm SF}$, very thin lines).  }
\label{tot} 
\end{flushleft}
\end{figure}

The evolution of Model 1 during the first 150 Myr is shown in
Fig.~\ref{9s}.  During the first few 10$^7$ yr, the central density is
not high enough to trigger SF; therefore, the system is only
characterised by a slow infall motion, due to the self-gravity.  After
$\sim$ 30 Myr the gas has been condensed enough to form the first
stars.  A network of cavities with increasing size and complexity is
created due to the patchy distribution of the SF sites.  It maintains,
however, a spherical symmetry and creates a gradually expanding
supershell.  The superbubble grows slowly in size, reaching $\sim$ 600
pc in 150 Myr.  Only at t $\sim$ 200 Myr does the size of the
superbubble reach 1 kpc, and at this point the SF rate (SFR) begins to
decrease, due to the reduction of available gas.  At the end of the
simulation, it is approximately one fourth of its peak value (see
Fig.~\ref{tot}).  The evolution of gas mass with time in the central
kpc of Model 1 (Fig.~\ref{tot}, third panel) is very similar to the
evolution of the SFR with time.  This is not a surprise, since we have
chosen a linear dependence between SF and the gas mass.  The increase
in the gas mass in the first $\sim$ 150 Myr is due to the fact that
the density distribution extends up to several kpc and the
self-gravity attracts a fraction of the gas located in the external
regions.  The global metallicity of the stellar component
(Fig.~\ref{tot}, lower panel) oscillates around a mean value of $\sim$
0.1 Z$_\odot$, consistent with what is observed in Fornax and in the
largest dSphs of the Local Group (Mateo \cite{mate98}).  For this
model we checked the timescale required to blow-away the gas from the
central kpc, and this corresponds to $\sim$ 400 Myr.  We reran this
model with higher spatial resolution, i.e. a resolution of 5 pc.  This
model shows a very similar behaviour to Model 1.

\begin{figure}[ht]
 \begin{flushleft}
 \vspace{-0.2cm}
 \hspace{0.8cm} \includegraphics[width=6.5cm]{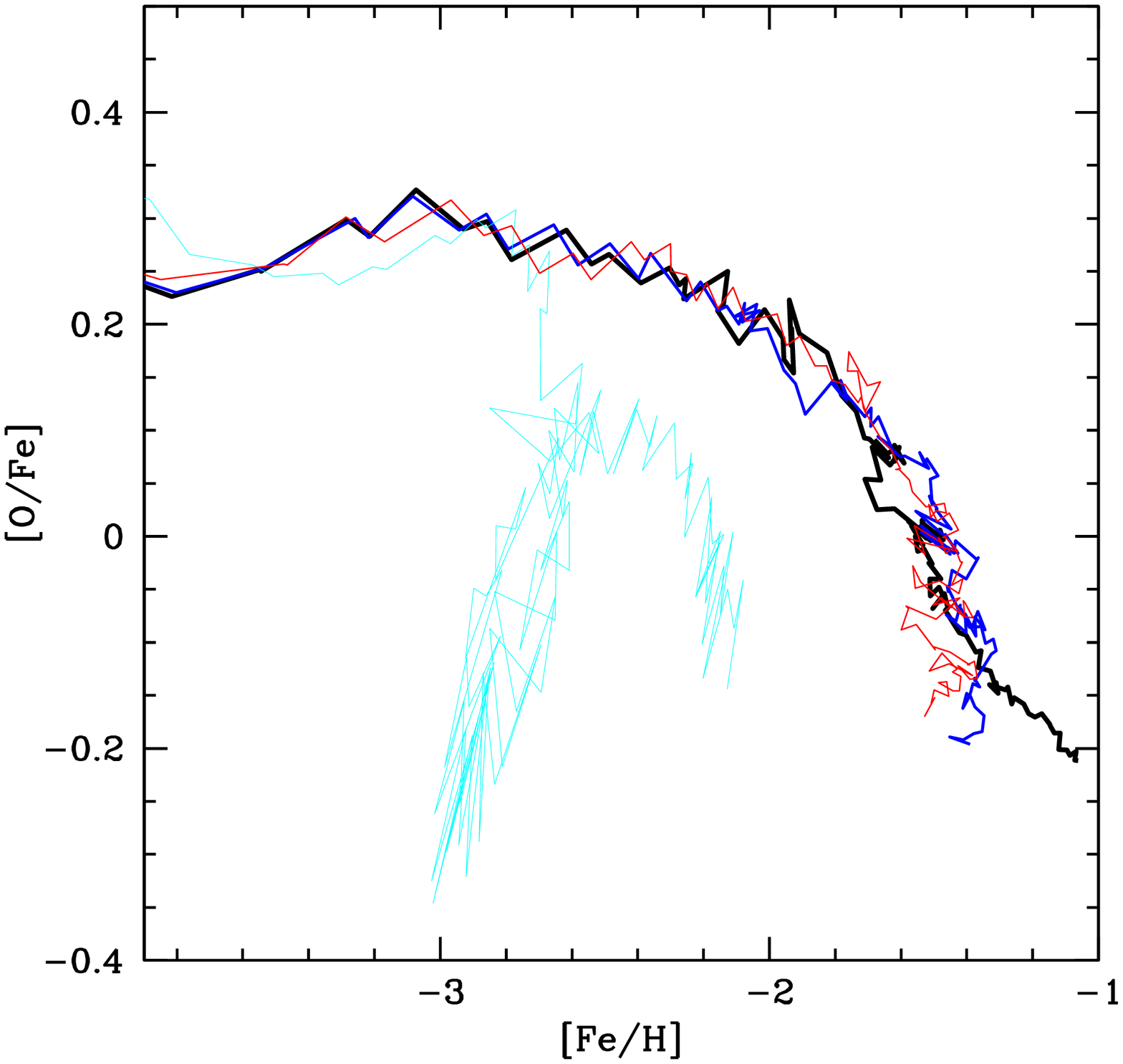}
\caption{ [O/Fe] vs. [Fe/H] in stars.  Symbols as in Fig.~\ref{tot}.}
\label{ofe} 
\end{flushleft}
\end{figure}

With our model, we are also able to check the build-up of the [O/Fe]
vs. [Fe/H] relation in the stellar population.  This is shown in
Fig.~\ref{ofe}.  The evolution in the first few 10$^7$ yr is dominated
by SNeII ejecta and this produces a plateau of [O/Fe].  At [Fe/H]
$\sim$ -2, the Fe produced by the SNeIa (which explode with some delay
compared to SNeII, Matteucci \& Greggio \cite{mg86}) starts reducing
the [O/Fe].  For Local Group dSphs, a knee at [Fe/H] $\sim$ -2 (much
earlier than the analogous knee observed in Milky Way stars) is a
common feature (see e.g. Tolstoy et al. \cite{tol03}; Lanfranchi \&
Matteucci \cite{lm04}).

\subsection{A parametric study}
\label{res_others}

With its lower temperature threshold for the onset of SF, Model 2 is
supposed to be characterised by a reduced impact of the feedback and
therefore a milder SFR due to the fact that a smaller amount of gas
fulfils the SF condition.  Indeed, its behaviour (Figs. \ref{tot} and
\ref{ofe}) is similar to the reference model, showing that SF
self-regulation (K\"oppen et al. \cite{kth95}) is at work, similar to
that in 1-D chemodynamical modelling of dwarf galaxies (Hensler et
al. \cite{htg04}).

In Model 3 we have suppressed the input of energy coming from stellar
winds, thereby reducing the feedback.  There is a significant effect
on the global thermal energy budget, resulting in a slightly larger
amount of gas inside the galaxy at the end of the simulation and,
consequently, in a larger SFR.  This model is more stable against the
feedback of the dying stars, so a longer duration of SF is expected.
At the end of the simulation, $\sim$ 10$^8$ M$_\odot$ of stars are
already present inside the galaxy (Fig. \ref{tot}), a higher value
than the estimated stellar mass of Fornax.  We have to take into
consideration, however, that more than 60 \% of these stars will die
within 10 Gyr.  Moreover, tidal stripping is likely to produce a
substantial reduction of the stellar component (Kroupa \cite{krou97}).

Model 4, with a reduced $\varepsilon_{\rm SF}$, has a higher initial
gas mass (see Sect. 2) and therefore starts forming stars earlier and
at a higher rate.  Its attained metallicity is of the order of a few
hundredth of Z$_\odot$, considerably lower than the observed value in
Fornax.  This is mostly due to how the newly produced metals are
diluted in a larger fraction of pristine gas but also to the fact that
the reduced injection energy rate per unit mass into the system
reduces the turbulence, therefore decreasing the diffusivity of the
medium and increasing the mixing timescale.  This value of the global
metallicity, although inconsistent with the chemical composition of
Fornax, is similar to the metallicity of smaller dSphs, supporting the
idea that the SF efficiency increases with the mass of the object
(e.g. Lanfranchi \& Matteucci \cite{lm04}).  The evolution of [O/Fe]
vs. [Fe/H] for this model shows a large loop (Fig. \ref{ofe}), caused
by the continuous increase in total gas mass through gas inflow
between $\sim$ 100 and $\sim$ 200 Myr that also produces a decrease in
the global metallicity (see Fig.~\ref{tot}).  Therefore, at variance
with the other models, the [Fe/H] is not a proxy of the time.

\begin{figure}[ht]
 \begin{flushleft}
 \vspace{-0.4cm}
 \hspace{0.4cm} \includegraphics[width=7cm]{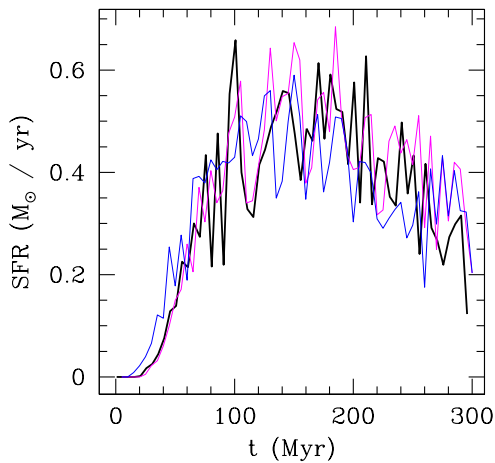}
 \vspace{-0.2cm}
\caption{ Evolution with time of the SFR (in M$_\odot$ / 
yr), for Model 1 (reference model, very thick line), Model 1de
(extended DM halo, thick line), and Model 1dc (concentrated DM halo, 
thin line).  }
\label{dm} 
\end{flushleft}
\end{figure}

Two DM-rich models (models 1dc and 1de) are also considered.
Their setup is identical to model 1, but their gravity is dominated
by a large static DM halo.  The comparison of the SFR in these models
is shown in Fig.~\ref{dm}.  The differences between the different
models are very small.  Also the other properties of the models are
not significantly affected by the presence of a large DM halo.  This
shows that the global behaviour of the gas in the central part of the
galaxy is mostly determined by the total mass and distribution of gas
itself rather than by the depth of the potential well.  However, it is
foreseeable that the late evolution of the gas is affected by the
presence of a DM halo extending up to very large distances.

We also ran a simulation in which $\varepsilon_{\rm SF}$=0.2.  In this
case, the expanding cavity created by overlapping SNe is energetic
enough to produce a blow-away and unbind most of the gas in the galaxy
in $\sim$ 100 Myr.  This determines the duration of the SF period,
which is very short and inconsistent with the typical SF histories of
Local Group dSphs (Grebel \cite{gre97}).

Finally, when an asphericity in the initial distribution of gas is
introduced, the gas tends to expand preferentially along one axis,
namely the one in which the pressure gradient is steeper.
Interestingly, the development of a polar galactic wind does not
suppress the SF process, since some gas is compressed towards the
walls of the galactic chimney where the condition for the onset of SF
can be fulfilled (see Recchi et al. \cite{rec07}).

\section{Discussion and conclusions}
\label{conc}

By means of a detailed 2-D chemodynamical code, we have simulated the
evolution of isolated dwarf spheroidal galaxies without DM in order to
study the early evolution of TDGs and compared the results with
DM-dominated models.

We have seen that reasonable assumptions about the SF efficiency and
the temperature threshold above which the SF is suppressed lead to a
SF lasting at least 400 Myr.  This interval of time broadens further
if we consider very small SF efficiencies (which is possible in small
galaxies).  Provided that it keeps its spherical symmetry, a small
galaxy is therefore relatively stable against the feedback of the
ongoing star formation, even in absence of a large DM halo.  This
issue has already been shown by Hensler et al. (\cite{htg04}).
However, we have seen that a DM-dominated model shows a behaviour
similar to the DM-free ones, indicating that the gas dynamics of the
galaxy, at least in its central part, is dominated by the gas density
and distribution more than by the depth of the potential well.
Eventually, after several hundred Myr, a strong galactic wind may
arise, able to eject most of the gas.  However, this gas does not
necessarily leave the parent galaxy.  It may indeed later be
reaccreted by the gravitational pull of the galaxy if tidal
interactions do not disperse it.  Recchi \& Hensler (\cite{rh06}) have
done experiments about the timescale required to refill with cold gas
the centre of a galaxy after a supernova-driven galactic wind has
occurred.  This timescale is of the order of a few hundred Myr, after
which a new episode of SF can occur.  This might lead to the bouncing
back and forth of the gas, due to the competing effects of the
gravitational pull and the feedback of the ongoing star formation,
resulting in a sort of self-regulation, until the moment in which a
close encounter with a more massive galaxy definitely moves the gas
away, shutting off the cycle.  This mechanism is under investigation
at the present time.

A more detailed description of the code and a more extensive analysis
of the parameter space will be provided in a paper in preparation.
Further work and modelling is necessary to sharpen our understanding
of TDGs.  In particular we are including stellar dynamics and a
time-dependent tidal field due to the interacting system.  Also the
IMF is going to be subject of detailed analysis.  In particular, we
will implement the integrated-galactic initial mass function (Weidner
\& Kroupa \cite{wk05}), in order to consider the variations of
galaxy-wide IMF as a function of the SFR.  Finally, we have started
3-D simulations of TDGs (Marcolini et al., in preparation).

\begin{acknowledgements}
  We thank the anonymous referee whose comments improved the paper.
  This project is supported by the German \emph{Deut\-sche
  For\-schungs\-ge\-mein\-schaft, DFG\/}, as part of the priority
  programme 1177 (under grants TH 511/8 and KR1635/8-1).  S.R. wishes
  to thank A. Rieschick for providing a subroutine for the
  calculation of the self-gravity and A. Marcolini for help and
  assistance in the development of the code.  
\end{acknowledgements}

\end{document}